\documentclass{article}
\pdfoutput=1
\usepackage{PRIMEarxiv}
\usepackage{tensor}
\newcommand{\dV}{
    d^4 x
}
\newcommand{\half}{
    \frac{1}{2}
}
\newcommand{\beq}{\begin{equation}}
        \newcommand{\eeq}{\end{equation}}
\newcommand{\beqs}{\begin{eqnarray}}
        \newcommand{\eeqs}{\end{eqnarray}}

\newcommand{\pd}[1]{\tensor{\partial}{#1}}

\newcommand{\covard}[1]{
    \tensor{\nabla}{#1}
}

\newcommand{\g}[1]{
    \tensor{g}{ #1 }
}

\newcommand{\riemann}[1]{
    \tensor{R}{ #1 }
}

\newcommand{\T}[1]{
    \tensor{T}{ #1 }
}

\newcommand{\Lagr}{
    \mathcal{L}
}

\newcommand{\B}[1]{
    \tensor{\mathcal{B}}{ #1 }
}
\newcommand{\G}[1]{
    \tensor{\mathcal{G}}{ #1 }
}

\usepackage[utf8]{inputenc} 
\usepackage[T1]{fontenc}    
\usepackage{hyperref}       
\usepackage{url}            
\usepackage{booktabs}       
\usepackage{amsfonts}       
\usepackage{nicefrac}       
\usepackage{microtype}      
\usepackage{lipsum}         
\usepackage{fancyhdr}       
\usepackage{graphicx}       
\usepackage{doi}
\usepackage{amsmath}
\usepackage{amssymb}
\usepackage{subfig}
\graphicspath{{plots/}}     

\renewcommand{\eqref}[1]{Eq.~(\ref{#1})}

\title{Orbital Motion of a test particle around a Schwarzschild's Black Hole in STVG gravity}

\author{
  {Devansh~Shukla} \\
    Sardar Vallabhbhai National Institute of Technology, Surat \\
    \texttt{i18ph021@phy.svnit.ac.in} \\
    \And
  {Abhay~Menon~A} \\
    Sardar Vallabhbhai National Institute of Technology, Surat \\
    \texttt{abhaymenon1996@gmail.com} \\
    \And
    {Kamlesh~Pathak} \\
    Sardar Vallabhbhai National Institute of Technology, Surat \\
    \texttt{knp@phy.svnit.ac.in} \\
}
\begin{document}
\maketitle
\begin{abstract}
  In this article, we have examined the existence of a static spherically symmetric solution in the Scalar Tensor Vector Gravity (STVG) and investigated its horizon distances to develop boundary limitations for our test particle. We have computed the Kretschmann invariant of the metric to study the singularities and verify that it reduces to general relativity's Kretschmann invariant as $\alpha\rightarrow0$. Further, we investigated the orbital motion of a time-like and light-like test particle around the static solution by developing an effective potential and the radius of the innermost stable circular orbit(ISCO).
\end{abstract}


\section{Introduction}
\paragraph*{}
The Scalar Tensor Vector Gravity theory, or STVG for short, is a recent modified theory of gravity developed by John W. Moffat in 2005.
The theory allows the gravitational constant $G$, a vector field coupling $\omega$ and the vector field mass $\mu$ to vary with spacetime by being scalar fields\cite{grqc_arxiv_1410_2464}.
The effect is that there is a Yukawa-like modification of the force, essentially predicting that the gravity is attractive and stronger than the Newtonian prediction at large distances and repulsive at small distances.

Although Einstein's General Relativity(GR) has been quite successful, it falls short in explaining the cosmic acceleration of the universe, galaxy rotation curves and galaxy velocities without the inclusion of the mysterious "dark" matter, which has to be about six times the baryonic matter to account for the disparity\cite{Zen_Vasconcellos_2020, arxiv.0711.1145}.
These arguments provide sufficient motivation to investigate modified gravity theories.
In this article, we will study one such modified theory called STVG, which has been able to explain the solar system observations\cite{grqc_arxiv_1410_2464}, the rotation curves of galaxies\cite{Moffat_2013, Moffat_2014} and the dynamics of galactic clusters\cite{Moffat_2015_rc} without the use of "dark" content.

Throughout the years, there has been excitement in the field of black hole physics in figuring out geodesics for test particles, whether they be neutral or charged, massive or massless, they help to study the nature of the spacetime in question. In order to investigate the geodesic motion, we utilize the parameter called innermost stable circular orbit(ISCO).
The ISCO can give us knowledge of the inner edge of the accretion disk, the spacetime geometry and the gravitational waves from the binary system\cite{1974ApJ191499P, wald2010general}.

As J.W. Moffat states in his paper\cite{Moffat_2006}, SVTG, nonsymmetric gravity theory(NGT), metric-skew-tensor gravity (MSTG)
have an essential feature that the modified acceleration law for weak gravitational fields has a repulsive Yukawa force, causing effective mass and coupling to matter to vary with distance rather than being attractive everywhere.
He computed the line elements for Schwarzschild, Reissner-Nordström, and the Kerr solution. In this article, we will utilize the line elements to compute and study the nature of the innermost stable circular orbits in the Schwarzschild's solution.

The paper is organized as follows. In section 2, we provide a brief explanation of the Scalar-Tensor-Vector gravity and its field equations. In section 3, we develop the static spherically symmetric solution's line element and examine its horizon distance(s). Next, we turn our attention to the innermost stable circular orbit for time-like and light-like particles in section 4. Finally, in section 5, we summarize the results.

We use the units $c = G_N = k_B = \hbar = 1$ and the sign convention $(-,+,+,+)$ throughout the paper. The metric in STVG assumes that $G = G_N(1 + \alpha)$, where $\alpha$ is a real non-negative constant and $G_N$ is the Newtonian gravitational constant.
\section{The ansatz model}\label{sec:ansatz_model}
\paragraph*{}
In order to obtain the ansatz model for STVG, we assume $G(x)$, $\mu(x)$ and $\omega(x)$ to be three scalar fields, with associated potential functions $V(G)$, $V(\mu)$ and $V(\omega)$ respectively.

Consider an anti-symmetric field $\B{_\mu _\nu}$ is formed out of the vector field $\phi_{\mu}$ such that:
\begin{equation}
  \B{_\mu _\nu} = \partial_\mu \phi_\mu - \partial_\nu\phi_\mu
\end{equation}
The total action, with $A_{M}$ as the matter action, reads:
\begin{equation}
  A = A_{g} + A_{\phi} + A_{S} + A_{M}  \label{eq:action}
\end{equation}
where
\begin{equation}
  A_{g} = \frac{1}{16\pi} \int \dV \sqrt{-g} \left[\dfrac{R}{G} + 2\Lambda\right]
\end{equation}
\begin{equation}
  A_{\phi} = - \int \dV \sqrt{-g} \omega \left[\dfrac{1}{4}\B{_\mu _\nu}\B{^\mu ^\nu} + V(\phi)\right]
\end{equation}
\begin{eqnarray}
  \begin{aligned}
    A_{S} = & \int \dV \sqrt{-g} \left[\dfrac{1}{G^3} \left(\half \g{^\mu ^\nu} \covard{_\mu} G \covard{_\nu} G - V(G) \right) + \dfrac{1}{G} \left(\half\g{^\mu ^\nu} \covard{_\mu} \omega \covard{_\nu} \omega - V(\omega) \right) \right. \\ & \left. + \dfrac{1}{\mu^2 G} \left(\half \g{^\mu ^\nu} \covard{_\mu}\mu \covard{_\nu}\mu - V(\mu)\right) \right]
  \end{aligned}
\end{eqnarray}
Here $\covard{_\mu}$ denotes the covariant derivate with respect to the metric $\g{_\mu _\nu}$

The total energy-momentum tensor is given by:
\begin{equation}
  \T{_\mu _\nu} = \T{_M _\mu _\nu} + \T{_\phi _\mu _\nu} + \T{_S _\mu _\nu}
\end{equation}
where $\T{_M _\mu _\nu}$ denotes the energy-momentum contribution by ordinary matter; $\T{_\phi _\mu _\nu}$ is the contribution from the vector field and $\T{_S _\mu _\nu}$ is by the scalar fields $G$, $\mu$ and $\omega$.
\begin{equation}
  \T{_M _\mu _\nu} = -\dfrac{2}{\sqrt{-g}} \dfrac{\delta A_{M}}{\delta \g{^\mu ^\nu}}; \quad \T{_\phi _\mu _\nu} = -\dfrac{2}{\sqrt{-g}} \dfrac{\delta A_{\phi}}{\delta \g{^\mu ^\nu}}; \quad \T{_S _\mu _\nu} = -\dfrac{2}{\sqrt{-g}} \dfrac{\delta A_{S}}{\delta \g{^\mu ^\nu}}
\end{equation}
We use the variational principle to compute the field equation by varying the \eqref{eq:action} against the metric $\g{^\mu ^\nu}$ to compute the field equation
\begin{equation}
  \G{_\mu _\nu} - \g{_\mu _\nu} \Lambda + Q_{\mu \nu} = 8\pi G \T{_\mu _\nu}  \label{eq:field_equation}
\end{equation}
where $Q_{\mu \nu} = G\left(\g{_\mu _\nu}\covard{^\alpha}\covard{_\alpha}\dfrac{1}{G(x)} - \covard{_\mu}\covard{_\nu}\dfrac{1}{G(x)}\right)$

This term is identical to the boundary term in the Brans-Dicke theory and STVG reduces to it under suitable conditions\cite{Brans_Dicke}.
\section{Static spherically symmetrical solution}
\paragraph*{}
In order to obtain the Schwarzschild's line element, we take an ansatz solution for the most general spherically symmetric metric:
\begin{equation}
  ds^2 = - Z(r) dt^2 + Z^{-1}(r) dr^2 + r^2 d\Omega^2
\end{equation}

J.W. Moffat\cite{Moffat_2015} computed the value of $Z(r_h)$ by assuming $\Lambda$ and $V(\phi)$ vanish while keeping the coupling constant $\omega=1$:
\begin{equation*}
  Z(r) = 1 - \dfrac{2GM}{r} + \dfrac{\alpha G_N G M^2}{r^2}
\end{equation*}
Since, $G_N = G(1 + \alpha)$, $Z(r)$ becomes:
\begin{equation}
  Z(r) = 1 - \dfrac{2(1 + \alpha)M}{r} + \dfrac{\alpha (1 + \alpha) M^2}{r^2}  \label{eq:sss_zr}
\end{equation}
The line element reads:
\begin{equation}
  ds^2 = - \left(1 - \dfrac{2(1 + \alpha)M}{r} + \dfrac{\alpha (1 + \alpha) M^2}{r^2}\right) dt^2 + \dfrac{dr^2}{\left(1 - \dfrac{2(1 + \alpha)M}{r} + \dfrac{\alpha (1 + \alpha) M^2}{r^2}\right)} + r^2 d\Omega^2  \label{eq:line_element}
\end{equation}
The Kretschmann invariant, defined as the contraction of the Riemannian tensor, $K = \riemann{_\alpha _\beta _\mu _\nu} \riemann{^\alpha ^\beta ^\mu ^\nu}$ reads:
\begin{equation}
  K = \dfrac{48M^2 (1 + \alpha)^2 r^2 - 96M^3 \alpha (1 + \alpha)^2 r + 56M^4 \alpha^2 (1 + \alpha)^2}{r^8}  \label{eq:sss_Kretschmann}
\end{equation}
The Kretschmann invariant reduces to general relativity's case for $\alpha \rightarrow 0$, and as $r \rightarrow 0$, the invariant becomes indeterminant, implying a singularity at $r=0$.

The modified line element \eqref{eq:line_element} will describe a black hole if there is a locatable horizon present where $Z(r)$ vanishes and $Z'(r_h)>0$ for positive surface gravity\cite{Calz2018}:
\begin{equation}
  r_{\pm} = M \left\{\left(1 + \alpha\right) \pm \sqrt{1 + \alpha}\right\}
\end{equation}
where $r_+$ and $r_-$ represents the outer(event horizon) and inner horizon(Cauchy horizon) respectively.

Let's consider the case where $\alpha = 0$, the horizons exists at $r_{+} = 2M$ and $r_{-} = 0$, which are identical to the general relativity's predictions.
\begin{figure}[h]
  \centering
  \subfloat[$M Z'(r_+)$ vs $\alpha$ \label{fig:stvg_hor_z_positive}]{\includegraphics[scale=0.42]{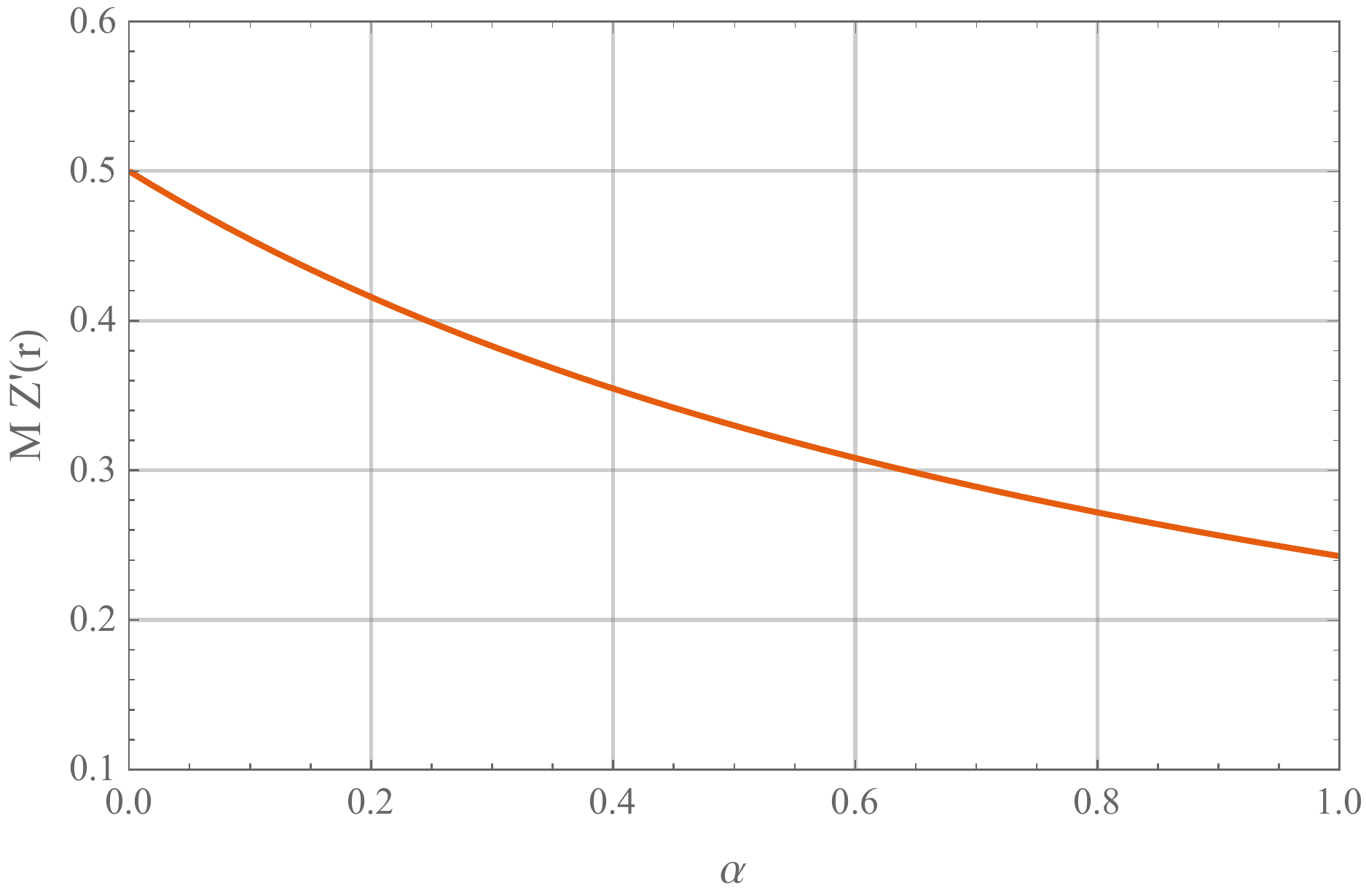}}
  \hfill
  \subfloat[$M Z'(r_-)$ vs $\alpha$ \label{fig:stvg_hor_z_negative}]{\includegraphics[scale=0.42]{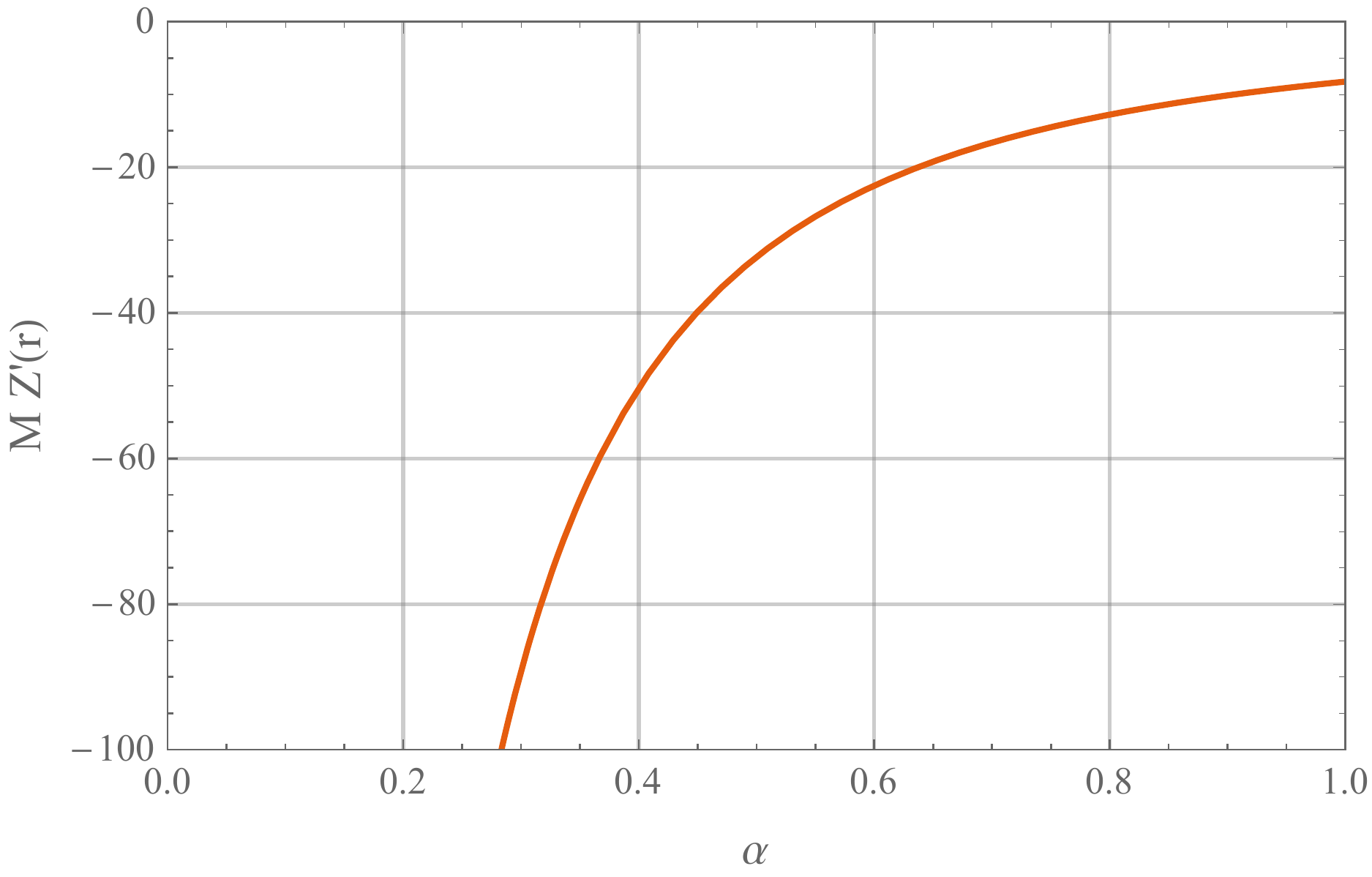}}

  \caption{The plots for $Z'(r)$ with respect to $\alpha$ for both solutions $r_\pm$}

  {\small $M Z'(r_+)$ remains positive against $\alpha$, indicating positive surface gravity}
\end{figure}

The static spherically symmetric solution, given by \eqref{eq:line_element}, describes a physical black hole because the root $r_+$ is real and positive while $M Z'(r_{+})$ remains positive against $\alpha$ (Fig. \ref{fig:stvg_hor_z_positive}).
\section{Test particle trajectory}
Let us assume a test particle of a non-negative mass $\mu$ in the equatorial frame ($\theta = \pi/2$). Its line element reads:
\begin{equation}
  ds^2 = \g{_\alpha _\beta} dx^\alpha dx^\beta = - Z(r) dt^2 + Z^{-1}(r) dr^2 + r^2 d\phi^2 = \epsilon  \label{eq:spacetime_kill_sub}
\end{equation}
where $\epsilon=-1,0,1$ for time-like, null-like, and space-like trajectories respectively

The Lagrangian of a free particle in a potential-less spacetime reads:
\begin{equation}
  \Lagr = \mu \sqrt{\g{_\alpha _\beta} \dot{x}^\alpha \dot{x}^\beta}  \label{eq:lagrangian_density}
\end{equation}
Since \eqref{eq:lagrangian_density} is independent of $t$ and $\phi$, the first integrals of motion obtained using the Euler-Lagrange equation reads:
\begin{equation}
  \mu Z(r) \dot{t} = E; \quad \mu r^2 \dot{\phi} = L  \label{eq:first_intergral_of_motion}
\end{equation}
where $E$ and $L$ represent the energy and the angular momentum of the test particle respectively.

Substituting \eqref{eq:first_intergral_of_motion} into the spacetime interval \eqref{eq:spacetime_kill_sub}:
\begin{equation}
  -\dfrac{E^2}{\mu^2 Z(r)} + \dfrac{\dot{r}^2}{Z(r)} + \dfrac{L^2}{\mu^2 r^2} = \epsilon
\end{equation}
\begin{equation}
  \dot{r}^2 + V^2_{\text{eff}} = \dfrac{E^2}{\mu^2}  \label{eq:orbit_solution}
\end{equation}
with the effective potential represented by $V_{\text{eff}} = \sqrt{\left(\dfrac{L^2}{\mu^2 r^2} - \epsilon\right)\left(1 - \dfrac{2GM}{r} + \dfrac{\alpha G_N M^2}{r^2}\right)}$

The extremum values of $V_{\text{eff}}$ determine the stable and unstable orbits of the test particle, with a circular orbit defined by $\dot{r} = 0$ and $V_{\text{eff},\; r} = 0$.

Solving for $\pd{_r}V_{\text{eff}} = 0$, we obtain:
\begin{equation}
  \dfrac{L^2}{\mu^2} = \frac{M r^2 \epsilon (1 + \alpha)(\alpha  M-r)}{r^2 - 3 (1 + \alpha) M r + 2 \alpha  (1 + \alpha) M^2}  \label{eq:angular_momentum_solution}
\end{equation}
Restricting ourselves to only circular orbits ($\dot{r}=0$), the energy equation reads:
\begin{equation}
  \dfrac{E^2}{\mu^2} =  -\frac{\epsilon  \left(r^2 - 2 (1 + \alpha) M r + \alpha  (1 + \alpha) M^2\right)^2}{r^2 \left(r^2 - 3 (1 + \alpha) M r + 2 \alpha  (1 + \alpha) M^2\right)}  \label{eq:energy_solution}
\end{equation}
\subsection{Time-like orbits: \texorpdfstring{$\epsilon=-1$}{ε=-1}}
\paragraph*{}
Choosing $\epsilon = -1$, we restrict ourselves to time-like orbits, with angular momentum and energy defined by \eqref{eq:angular_momentum_solution} and \eqref{eq:energy_solution}.

\begin{figure}[h]
  \centering
  \includegraphics[scale=0.65]{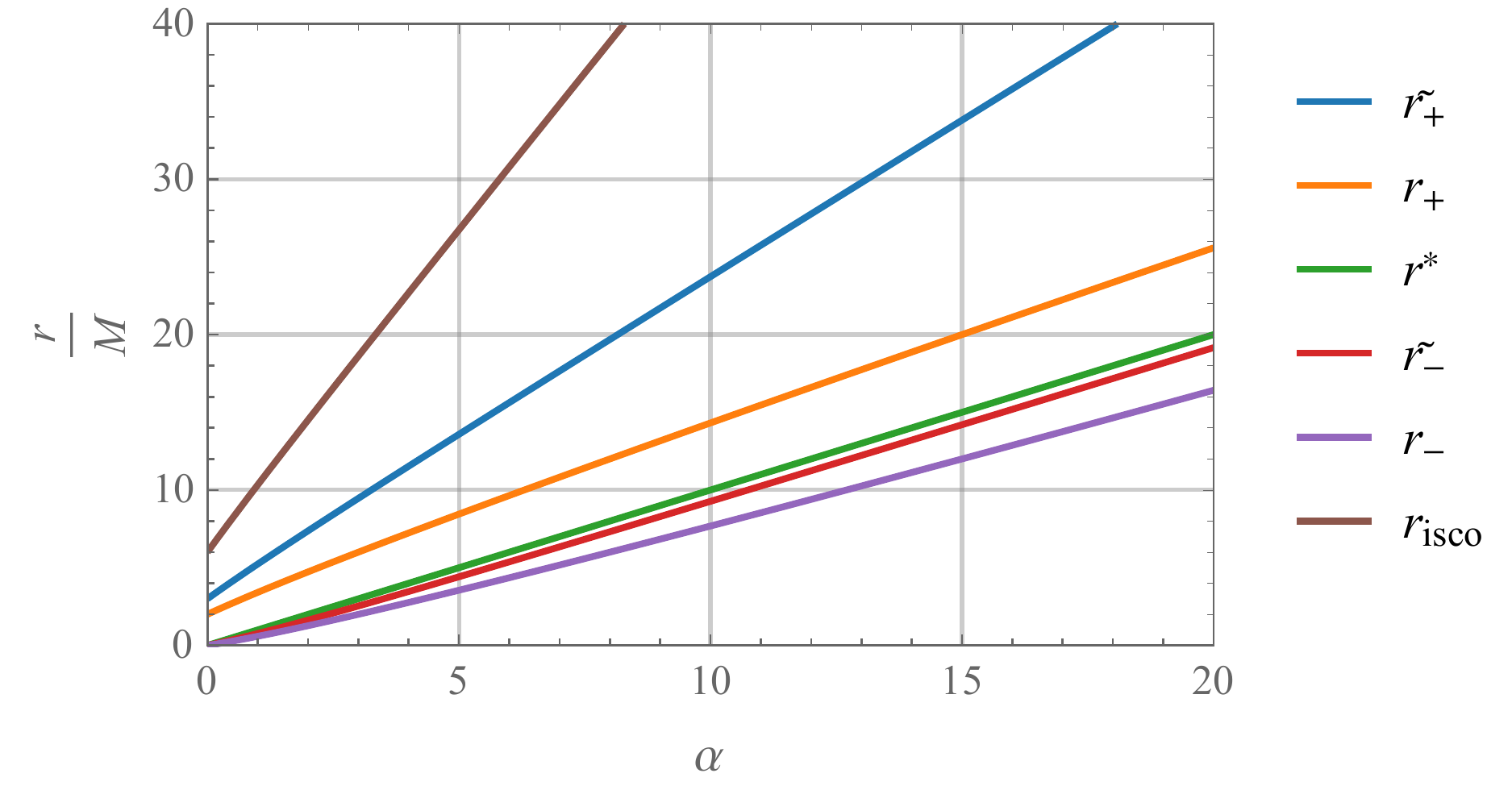}
  \caption{Plot of $\widetilde{r}_{\pm}$, $r_{\pm}$, $r^*$ and $r_{isco}$ per unit $M$ vs $\alpha$ \label{fig:r_vs_alpha}}

  {At $\alpha = 0$, $r_{isco} = 6M$, $\widetilde{r}_{+} = 3M$, $r_{+} = 2M$ and $\widetilde{r}_{-} = r_{-} = r^* = 0$ }
\end{figure}

The existence of a real-valued angular momentum implies two additional conditions for the existence of motion.
\begin{equation}
  r > \alpha M = r^*; \quad r^2 - 3 (1 + \alpha) M r + 2 \alpha  (1 + \alpha) M^2 > 0  \label{eq:two_additional_roots}
\end{equation}
Solving \eqref{eq:two_additional_roots}, we obtain two additional roots:
\begin{equation}
  \widetilde{r}_{\pm} = \frac{M}{2} \left(3 (1 + \alpha) \pm\sqrt{\alpha ^2+10 \alpha +9}\right)
\end{equation}
Implying from Fig. \ref{fig:r_vs_alpha}, we can see that the $r$'s follow linear functions with respect to $\alpha$, with $r_-$ and $\widetilde{r}_+$ as the extremum values:
\begin{equation}
  r_- < \widetilde{r}_{-} < r^* < r_+ < \widetilde{r}_{+}
\end{equation}
In the case of time-like circular orbits, by choosing the extremum value, the motion occurs in the region $r>\widetilde{r}_+$.

To obtain the radius of the circular orbit, we substitute the value of the angular momentum \eqref{eq:angular_momentum_solution} into the effective potential and get its extremum values by solving for $\partial_{r} V_{\text{eff}}(r) = 0$:
\begin{equation}
  \left(r^2 + G M (\alpha  G_N M-2 r)\right) \left(r^3-G M \left(4 \alpha ^2 G_N^2 M^2-9 \alpha  G_N M r+6 r^2\right)\right) = 0
\end{equation}
\begin{eqnarray}
  \begin{aligned}
    \dfrac{r_{isco}}{M} = 2 (1 + \alpha) & + \left({\alpha\left(\alpha + 1 \right)\sqrt{(\alpha + 5)}+\left(\alpha ^3+8 \alpha ^2+15 \alpha +8\right)}\right)^{1/3} \\ & + \frac{\alpha ^2+5 \alpha + 4}{\left({\alpha\left(\alpha + 1 \right)\sqrt{(\alpha +5)}+\left(\alpha ^3+8 \alpha ^2+15 \alpha +8\right)}\right)^{1/3}}
  \end{aligned}  \label{eq:risco}
\end{eqnarray}
Evaluating \eqref{eq:risco} at $\alpha = 0$, we obtain $r_{isco} = 6 M$, which agrees with the predictions of general relativity and as expected $r_{isco} > \widetilde{r}_+$, as shown in Fig. \ref{fig:r_vs_alpha}.
\begin{figure}[h]
  \subfloat[Energy per unit mass as a function of $\dfrac{r}{M}$(for $\dfrac{r}{M}>1$)]{\includegraphics[scale=0.42]{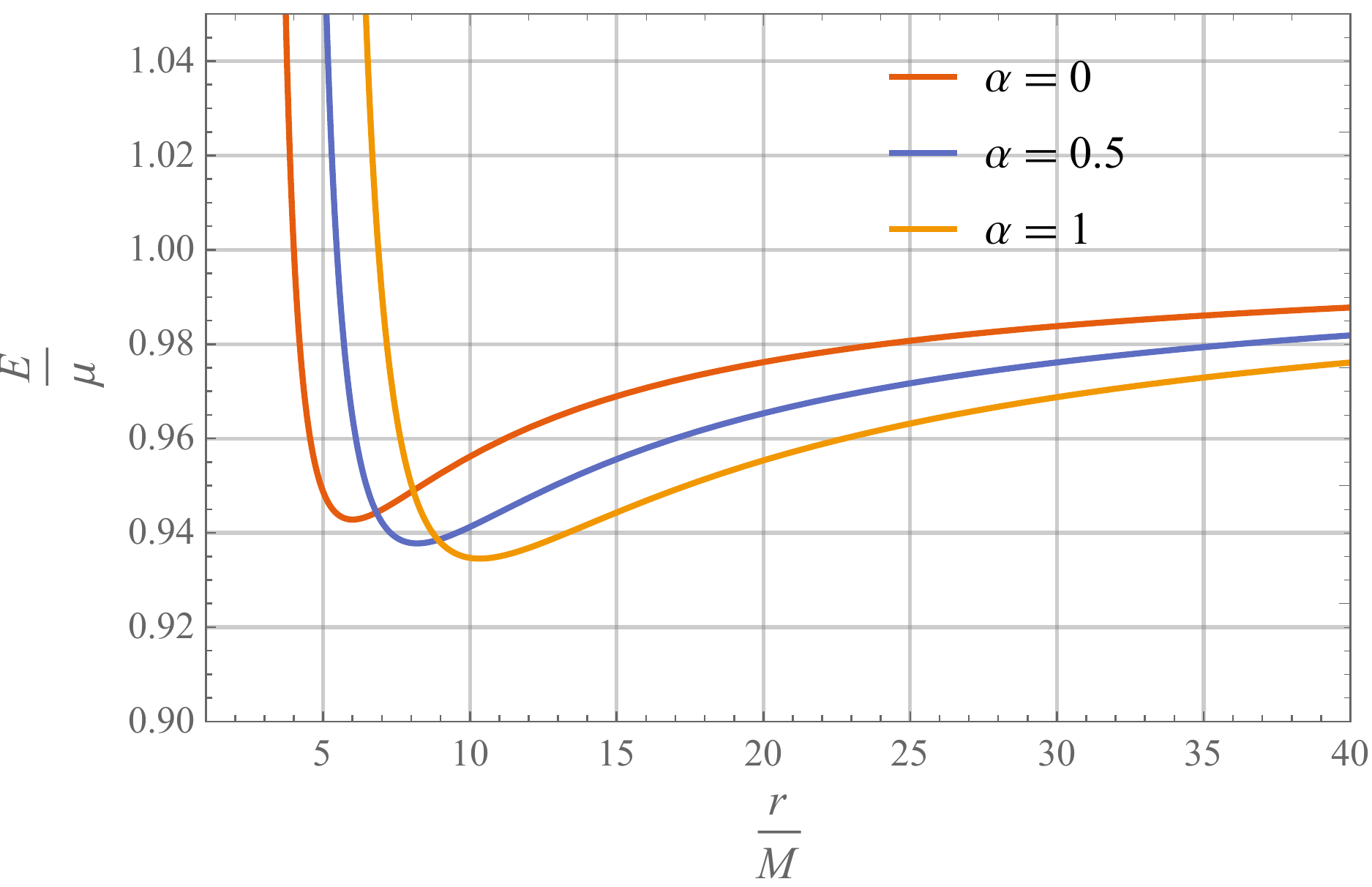}\label{fig:energy_stvg_timelike}}
  \hfill
  \subfloat[Angular momentum per unit mass as a function of $\dfrac{r}{M}$]{\includegraphics[scale=0.42]{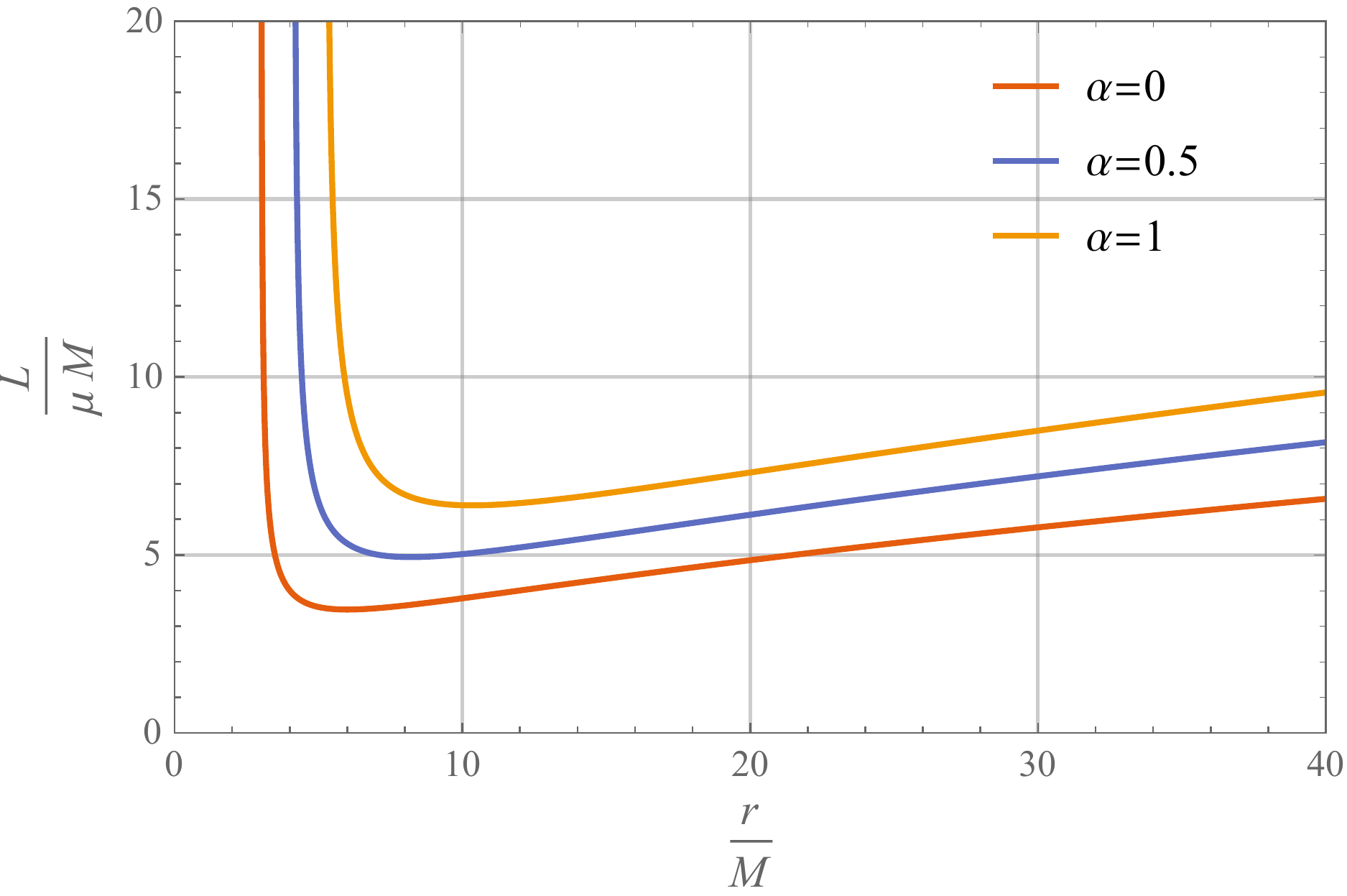}}
  \caption{Plot of (a) $\dfrac{E}{\mu}$ and (b) $\dfrac{L}{\mu M}$ as a function of $\dfrac{r}{M}$, for $\alpha=0, 0.5, 1$\label{fig:energy_momentum_timelike}}
\end{figure}
\subsection{Light-like orbits: \texorpdfstring{$\epsilon=0$}{ε=0}}
Choosing $\epsilon=0$, we restrict ourselves to light-like particles(null trajectories).

Differentiating the effective potential and computing its extremum values while assuming a non-zero angular momentum, we obtain:
\begin{equation}
  r^2 + (\alpha +1) M (2 \alpha  M-3 r) = 0  \label{eq:null_risco_root}
\end{equation}
The real and positive roots of \eqref{eq:null_risco_root} will define the innermost stable orbits for light-like particles:
\begin{equation}
  r_{isco}^{\pm} = \frac{M}{2} \left(3(\alpha + 1) \pm \sqrt{\alpha ^2+10 \alpha +9}\right)  \label{eq:null_risco}
\end{equation}
Evaluating \eqref{eq:null_risco} at $\alpha = 0$, we obtain the $r^+_{isco} = 3M$ and $r^-_{isco} = 0$, we can safely neglect the $r^-_{isco}$ since no motion is not possible below $\widetilde{r}_+$.
Solely focusing on $r_{isco} = r^+_{isco} = 3M$, we can see it is just as predicted by general relativity.
\begin{figure}[h]
  \centering
  \includegraphics[scale=0.45]{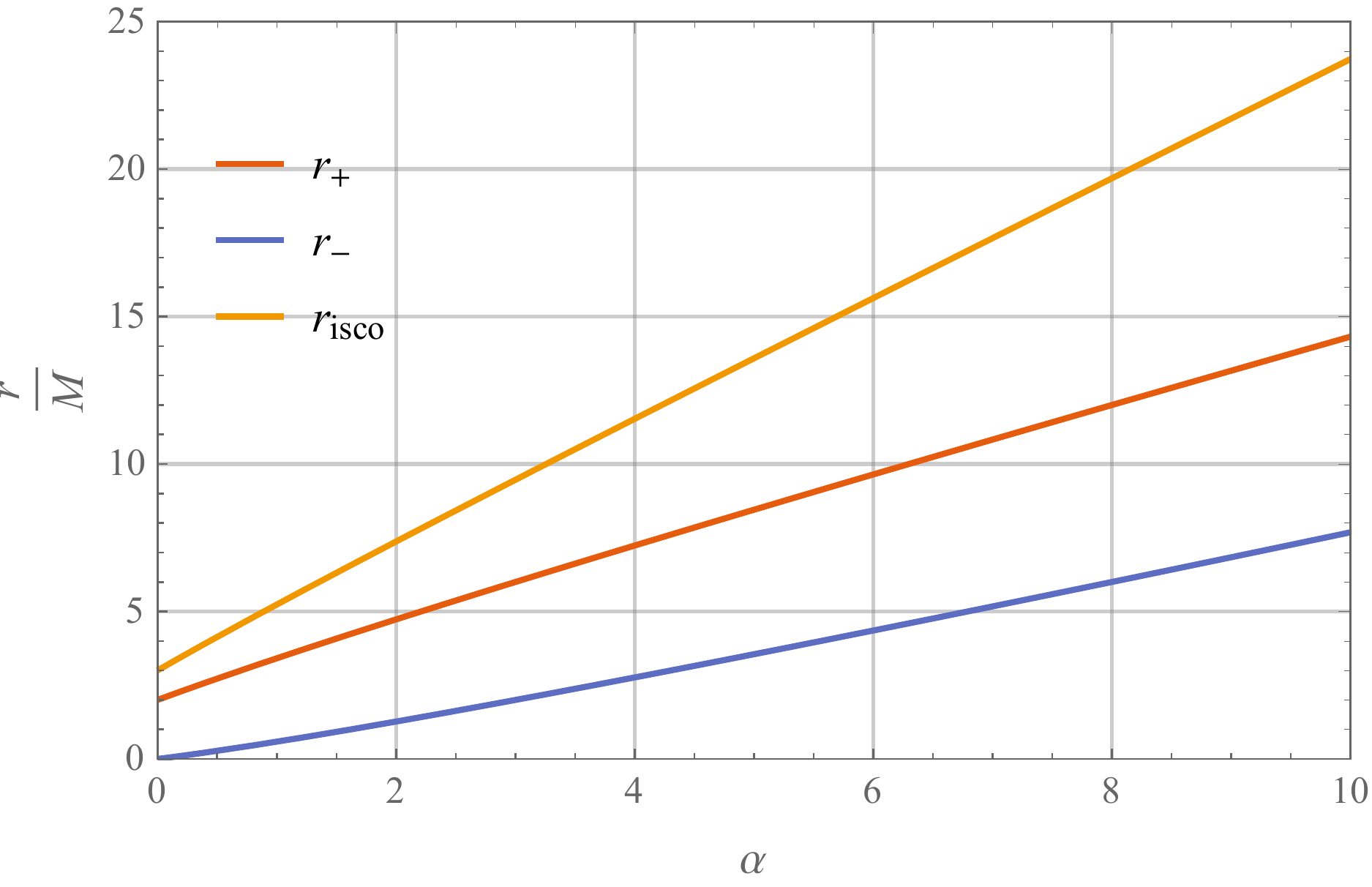}
  \caption{Plot of light-like $r_{+}$, $r_{-}$ and $r_{isco}$ per unit $M$ as a function of $\alpha$  \label{fig:light_r_plot}}

  {At $\alpha=0$, $r_{isco}=3M$, $r_{+}=2M$, $r_{-}=0$}
\end{figure}

As shown in Fig. \ref{fig:light_r_plot}, the $r_{isco}$, $r_+$ and $r_-$ are all proportional to $\alpha$ and $r_{isco}>r_+>r_-$
\section{Conclusion}
\paragraph*{}
In this article, we have developed the Schwarzschild's line element in the Scalar-Tensor-Vector gravity and examined it's Kretschmann's invariant, given by \eqref{eq:sss_Kretschmann}.
The Kretschmann's invariant is indeterminable at $r=0$ implying a singularity.
Analyzing the horizon distances, we obtain two values: the event horizon ($r_+$) and the Cauchy horizon($r_-$).
As expected, the Cauchy horizon is unstable, similar to the case of the Reissner-Nordström solution.

The orbit equation of a neutral particle around the black hole solution is given by \eqref{eq:orbit_solution} with the angular momentum defined by \eqref{eq:angular_momentum_solution} and the energy for the circular orbit by \eqref{eq:energy_solution}.
We study the nature of energy momentum and energy for time-like particles in Fig. \ref{fig:energy_momentum_timelike} for multiple values of $\alpha$.

We see that for $\epsilon = 0$(null-like) and $\epsilon=-1$(time-like), the innermost stable orbit is a strictly increasing function of $\alpha$. This means that, in both cases, we see an outward shift of $r_{isco}$ as the strength of the gravitational field increases.
This is expected since stronger gravitational fields cause greater perturbations closer to the source. We notice that as $\alpha \rightarrow 0$, the $r_{isco}$ reaches the GR predictions of $6M$ and $3M$ for time-like and null-like trajectories, respectively.

\nocite{*}
\bibliographystyle{unsrt}
\bibliography{citations}
\end{document}